\newcommand{\mrm}{\mathrm}
\newcommand{\pt}{\ensuremath{p_{\mathrm{t}}}}
\newcommand{\dd}{\mrm{d}}

\documentclass[epj]{svjour}
\usepackage{graphics}
\begin{document}
\title{Anisotropic Flow from RHIC to the LHC}
\author{Raimond Snellings for the STAR and ALICE Collaborations
}                     
\offprints{Raimond.Snellings@nikhef.nl}   
\institute{NIKHEF, Kruislaan 409, 1098 SJ, Amsterdam, The Netherlands}
\date{Received: date / Revised version: date}
%
\abstract{
Anisotropic flow is recognized as one of the main observables providing 
information on the early stage of a heavy-ion collision.
At RHIC the large observed anisotropic flow 
and its successful description by ideal hydrodynamics is considered evidence 
for an early onset of thermalization and almost ideal fluid properties of the 
produced strongly coupled Quark Gluon Plasma. 
This write-up discusses some key RHIC anisotropic flow measurements 
and for anisotropic flow at the LHC some predictions.
} 
\maketitle
%
\section{Introduction}
\label{sec:intro}

Flow is an ever-present phenomenon in nucleus--nucleus collisions, from
low-energy fixed-target reactions 
up to $\sqrt{s_{_{\rm NN}}}=200$~GeV collisions at the 
Relativistic Heavy Ion Collider (RHIC), and is expected
to be observed at the Large Hadron Collider (LHC).  
Flow signals the presence of multiple interactions
between the constituents
and is an unavoidable consequence of thermalization.

The usual theoretical tools to describe flow are
hydrodynamic or microscopic transport (cascade) calculations.  Flow
depends in the transport models on the opacity, be it
partonic or hadronic.  Hydrodynamics becomes valid 
when the mean free path of particles is much smaller than the system
size and allows for a description of the system in terms of macroscopic
quantities. This gives a handle on the equation of state of the
flowing matter and, in particular, on the value of the sound
velocity~\cite{Ollitrault:1992bk}.  In both types of models it may be possible
to deduce from a flow measurement whether the flow originates from partonic
or hadronic matter or from the hadronization
process~\cite{Danielewicz:1999vh,Rischke:1996nq,Ollitrault:1997vz}.

A convenient way of characterizing the various patterns of anisotropic
flow is to use a Fourier expansion of the triple differential
invariant distributions~\cite{Voloshin:1994mz}:
\[
E \frac{\dd^3 N}{\dd^3 {\bf p}} = 
\frac{1}{2\pi} \frac {\dd^2N}{\pt\,\dd\pt\,\dd y} 
  \left\{ 1 + 2\sum_{n=1}^{+\infty}  v_n \cos[n(\varphi-\Psi_R)]
  \right\}, 
\label{FLOW:ed3t}
\]
where $\varphi$ and $\Psi_R$ are the particle and reaction-plane azimuths
in the laboratory frame, respectively.  The sine terms in such an
expansion vanish due to reflection symmetry with respect to the
reaction plane. The Fourier coefficients are given by
\[
\label{FLOW:defvn}
v_n(\pt,y) = \langle \cos[n(\varphi-\Psi_R)] \rangle,
\]
where the angular brackets denote an average over the 
particles, summed over all events, in the $(\pt,y)$ bin under study.  
In this parameterization, the first two coefficients, $v_1$ and $v_2$, 
are known as directed and elliptic flow, respectively.

\section{Elliptic Flow: $v_2$}
\label{sec:v2rhic}

\begin{figure}[b]
\resizebox{0.48\textwidth}{!}{
  \includegraphics{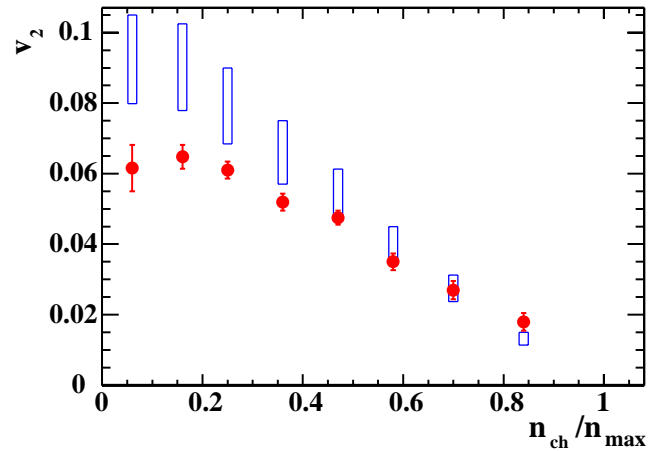}
}
\caption{ Elliptic flow (solid points) as a function of centrality
  defined as $n_{ch}/n_{max}$. The open rectangles show a range of
  values expected for $v_2$ in the hydrodynamic limit, scaled from
  $\epsilon$, the initial space eccentricity of the overlap
  region. 
  From~\cite{Ackermann:2000tr}}
\label{fig:v2cen} 
\end{figure}
Elliptic flow has its origin in the amount of rescattering and the
spatial eccentricity of the collision zone. 
The amount of rescattering is expected to increase with increasing centrality, 
while the spatial eccentricity decreases.  
This combination of trends dominates the
centrality dependence of elliptic flow.
The spatial eccentricity is defined by
\[
\epsilon = \frac{\langle y^2 - x^2 \rangle}{\langle y^2 + x^2 \rangle}
\]
where $x$ and $y$ are the spatial coordinates in the plane
perpendicular to the collision axis. The brackets $\langle \rangle$
denote an average weighted with the initial density.

Figure~\ref{fig:v2cen} shows the first measurement of elliptic 
flow at RHIC~\cite{Ackermann:2000tr}. 
Generally speaking, large values of elliptic flow are considered 
signs of hydrodynamic behavior as was first put forward by
Ollitrault~\cite{Ollitrault:1992bk}.
In hydrodynamics $v_2$ is essentially proportional to the spatial eccentricity 
(the strength depends on the velocity of sound of the matter).
The open rectangles in Fig.~\ref{fig:v2cen} show, for a range of 
possible values of the velocity of sound, the expected $v_2$ values 
from ideal hydrodynamics.
For ~$n_{ch}/n_{max} \ge 0.5$ (${\bf b} \le 7$ fm) it is observed 
that the data is well described by ideal hydrodynamics. 

The observed large amount of collective flow, 
in particular elliptic flow, 
is one of the main experimental discoveries at 
RHIC~\cite{Arsene:2004fa,Back:2004je,Adams:2005dq,Adcox:2004mh} and the 
main evidence suggesting nearly perfect fluid 
properties of the created matter~\cite{RHIC_Discoveries}.

\begin{figure}[thb]
\resizebox{0.48\textwidth}{!}{
  \includegraphics{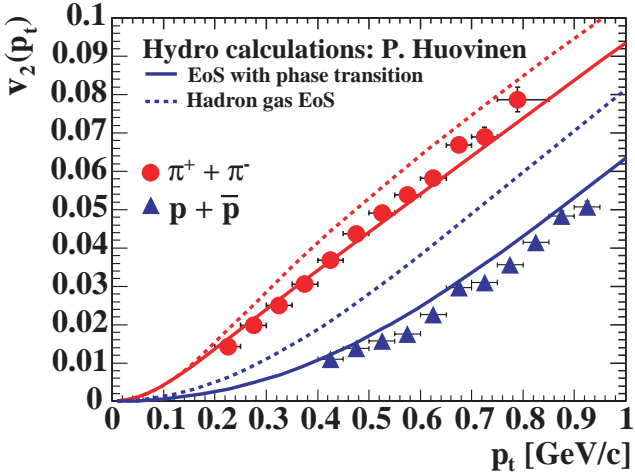}
}
\caption{ 
Elliptic flow of pions and protons as function of transverse 
momentum~\cite{Adler:2001nb}.
The lines are hydrodynamical model calculations using two different 
Equations of State (EoS), the dashed lines represent calculations done 
with a hadron gas EoS while the solid curves are calculation with an EoS 
which incorporates the QCD phase transition.
} 
\label{fig:v2pid} 
\end{figure}
Figure~\ref{fig:v2pid} shows $v_2$ for identified
particles as function of transverse momentum. 
At low $\pt$ the elliptic flow depends on the mass of the particle with 
$v_2$ at a fixed $\pt$ decreasing with increasing mass.
This dependence is expected in a scenario where all the particles 
have a common radial flow velocity~\cite{Borghini:2001ys,Blaizot:2001tf} 
as shown by the curves in Fig.~\ref{fig:v2pid} from ideal hydrodynamics.
The difference between the dashed and solid curves is the EoS, the dashed
curves correspond to calculations done with a hadron resonance gas EoS 
while the solid curves are hydro calculations incorporating the QCD 
phase transition.
The sensitivity to the EoS is better for the heavier particles because 
they are less affected by the contribution of the finite freeze-out 
temperature. 
It is clear that the hydro calculations incorporating the 
QCD phase transition give better description of the observed mass 
splitting.
However, detailed constrains on the EoS can only be obtained with better
modeling of the hadronic 
phase~\cite{Teaney:2001av,Teaney:2000cw,Hirano:2005wx,Hirano:2005xf} 
and the transition~\cite{Huovinen:2005gy} between the QGP and 
hadronic phase.

\begin{figure}[hbtp]
\resizebox{0.48\textwidth}{!}
{
\includegraphics{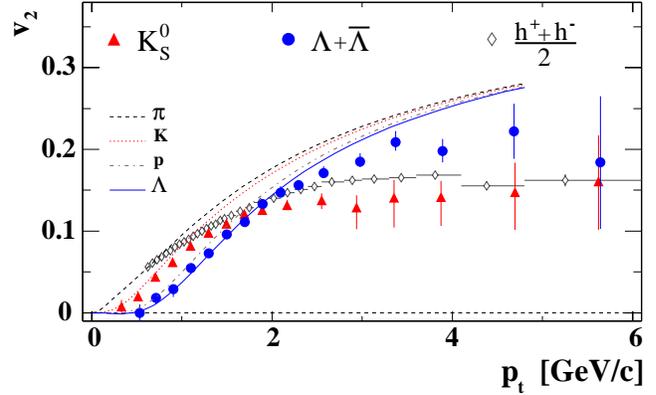}
}
\caption{
  The minimum-bias (0--80\% of the collision
  cross section) $v_{2}(p_T)$ for $K_{S}^{0}$,
  $\Lambda+\overline{\Lambda}$ and $h^{\pm}$. 
  Hydrodynamical calculations of $v_2$ for pions,
  kaons, protons and lambdas are also
  plotted~\cite{Huovinen:2001cy}. From~\cite{Adams:2003am}
} 
\label{fig:kaonlambda}
\end{figure}
In ideal hydrodynamics the mass ordering in $v_2$ persists up to arbitrary 
large $\pt$, although less pronounced because the $v_2$  
of the different particles start to approach each other.
Figure~\ref{fig:kaonlambda} shows that at higher $\pt$ the measurements 
start to deviate
significantly from hydrodynamics for all particle species,
and that the observed $v_2$ of the heavier 
baryons is larger than that of the lighter mesons. This mass
dependence is the reverse of the behavior observed at low $\pt$.
This is not expected in hydrodynamics and is also not expected if the 
$v_2$ is caused by parton energy loss (in the latter case 
there would, to first order, be no particle type dependence).
An elegant explanation of the unexpected particle type dependence
and magnitude of $v_2$ at large $\pt$ is provided by the coalescence 
picture~\cite{Voloshin:2002wa,Molnar:2003ff}. 

\begin{figure}[htb]
\resizebox{0.48\textwidth}{!}
{
  \includegraphics{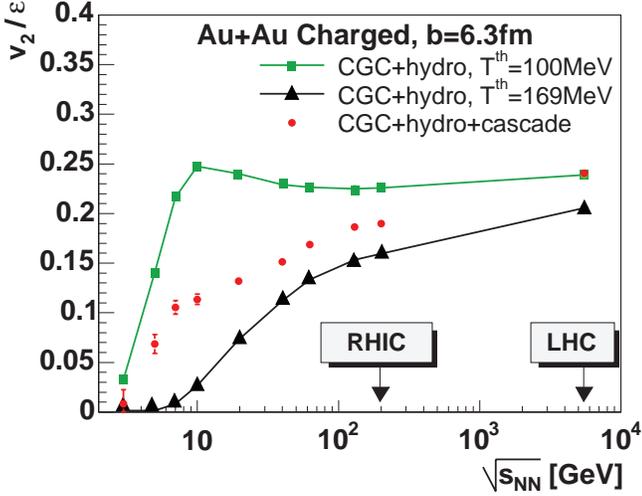}
}
\caption{
  Theoretical predictions~\cite{Hirano:2005} of $v_2/\epsilon$ versus 
  collision energy using color glass condensate estimates for the initial
  conditions. 
  Ideal hydrodynamic expansion up to kinetic freeze-out (squares) or chemical
  freeze-out (triangles) is assumed. 
  The full circles are results using a
  hadronic cascade model to describe the final phase after chemical
  freeze-out. 
}
\label{fig:v2LHC} 
\end{figure}

With the models which successfully describe the properties of the 
matter created 
at RHIC one can (and should) make predictions for the LHC. 
Testing these predictions will provide important confirmation of, 
or perhaps new 
insights to, our current understanding of QCD matter.
Figure~\ref{fig:v2LHC} shows elliptic flow calculations for the LHC. 
Using Color Glass Condensate (CGC) estimates for the initial condition 
the flow is calculated using ideal hydrodynamics up to the kinematic freeze-out
temperature of 100~MeV (full squares and upper curve). More realistic
estimates are obtained by assuming hydrodynamics up to the chemical
freeze-out temperature of 169~MeV followed by a hadron cascade
description of the final phase (full circles). The contribution from
the QGP phase ({\it i.e.} hydrodynamics up to 169~MeV) is shown by the
triangles (and lower curve) in the figure. 
It is seen from Fig.~\ref{fig:v2LHC} that at LHC energies 
the contribution from
the QGP phase is much larger than at RHIC or SPS, and that, as a 
consequence, there is less uncertainty due to the detailed modeling of
the hadronic phase. 
Theoretical calculations
such as these or {\it e.g.} in Ref.~\cite{Kolb:2000sd}, as well as
straight-forward extrapolations from lower energies based on particle
multiplicities predict maximum flow values of about 5--10\% at the LHC.
If the flow values (and corresponding multiplicities) at the LHC are 
indeed that large then the flow measurement should be relatively easy.

However, the previous hydro estimates assume that during 
the QGP phase the matter has zero shear viscosity. 
Teaney~\cite{Teaney:2003kp} has shown that even a small shear viscosity 
has a large effect on the buildup of flow. 
Recent calculations~\cite{Hirano:2005wx,Csernai:2006zz} show how the 
viscosity increases from RHIC to the LHC. To estimate how 
this would affect the predicted flow, viscous corrections have to be 
implemented in hydro models~\cite{Heinz:2005zi}.

In addition, experimental measurements of flow are affected by biases 
from physical effects unrelated to anisotropic flow (`non-flow
effects' like e.g. jet correlations) or due to additional features of 
the flow signal itself (e.g. 
fluctuations~\cite{Miller:2003kd,Voloshin:2006gz,Bhalerao:2006tp,Alver:2006pn}).
To estimate the effect of jet like correlations at the LHC a simple 
estimate can be made similar to what was done for the first RHIC flow 
measurement~\cite{Ackermann:2000tr}. 
The estimate of the non-flow is given by
\begin{equation}
\label{eq:nonflow}
\langle \cos[n(\Psi_2^a-\Psi_2^b)] \rangle \propto M_{sub} v_2^2  
+\tilde{g},
\end{equation}
where the angular brackets denote an average over the events, 
$\Psi_2^{a,b}$ are the subevent event planes, $M_{sub}$ is the 
corresponding subevent multiplicity and $\tilde{g}$ is the non-flow 
component.
The estimated value of $\tilde{g}$ from HIJING at 
$\sqrt{s_{NN}} =$ 130 GeV in the STAR 
acceptance using random subevents was 0.05. 
In the ALICE TPC acceptance 
the value of $\tilde{g}$ from HIJING at $\sqrt{s_{NN}} =$ 5.5 TeV is 
found to be 0.08. With better tuned definitions of the subevents 
the value of $\tilde{g}$ could easily be reduced to 0.04.
The correlation due to flow, $M_{sub} v_2^2$, is expected to be much 
larger than the non-flow contribution, 0.04, in a large centrality range 
for Pb+Pb collisions measured by ALICE at the LHC. 
This then indeed suggest that 
measuring flow can be done in great detail at the LHC.

\section{Higher Harmonics}
\label{sec:v4}

Higher harmonics of the momentum anisotropy are generally expected to be
small~\cite{Kolb:1999it,Teaney:1999gr}. More recently it
was realized that at higher $\pt$ they may become significant, 
in addition, that they are sensitive to the initial 
conditions~\cite{Kolb:2003zi} and that they depend on the equation of
state~\cite{Huovinen:2005gy}. 
In Ref.~\cite{Borghini:2005kd,Bhalerao:2005mm} it is argued
that in particular $v_4$ in combination with $v_2$ probes 
ideal fluid behavior, because for an ideal fluid the ratio 
$v_4(\pt)/v_2^2(\pt)$ should approach 0.5.  

\begin{figure}[htb]
\resizebox{0.48\textwidth}{!}{
  \includegraphics{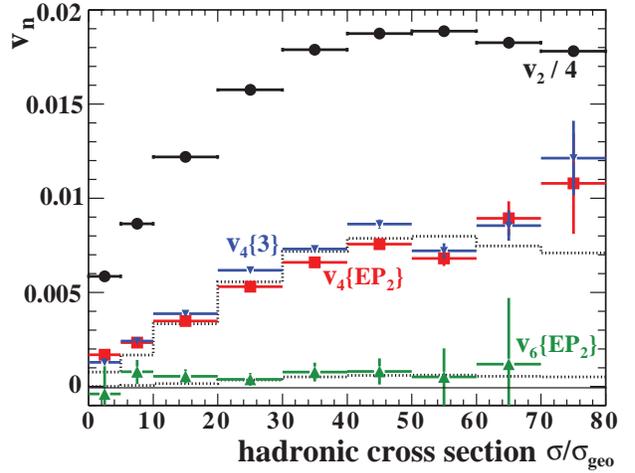}
}
\caption{
 The $p_t$- and $\eta$- integrated values of $v_2$,
$v_4$, and $v_6$ as a function of centrality. 
The dotted histograms are $1.4 \cdot v_2^2$ and $1.4 \cdot v_2^3$. 
In $v_n\{\}$ the term in the curly brackets indicates the method 
used~\cite{Borghini:2001vi}.
Figure from~\cite{Adams:2003zg}.
}
\label{fig:vncen} 
\end{figure}
The measured $\pt$-integrated $v_4$ and $v_6$ as function of 
centrality are shown in Fig.~\ref{fig:vncen}~\cite{Adams:2003zg}.
For comparison, $v_2$ is shown in the same figure. The integrated
$v_4$ is an order of magnitude smaller than $v_2$, as expected. The
higher harmonics $v_6$ and $v_8$ (not shown) are consistent with zero.
Figure~\ref{fig:vncen} shows that the ratio $v_4/v_2^2$ is larger 
than unity for all centralities which seems in contradiction with the 
prediction for ideal fluid behavior. 
However Ref.~\cite{Borghini:2005kd,Bhalerao:2005mm} shows that this 
asymptotic value of $v_4 = v_2^2/2$ is reached at transverse momenta 
well above 1 GeV/c. 
At these higher transverse momenta hydrodynamics is expected 
to break down and thus the ratio $v_4/v_2^2$ expected to increase again. 
In addition, at RHIC the integrated ratio of $v_4/v_2^2$ is mainly 
determined by particles below 1 GeV/$c$, and is therefore not so well 
suited for this comparison.  

\begin{figure}
\resizebox{0.48\textwidth}{!}{
  \includegraphics{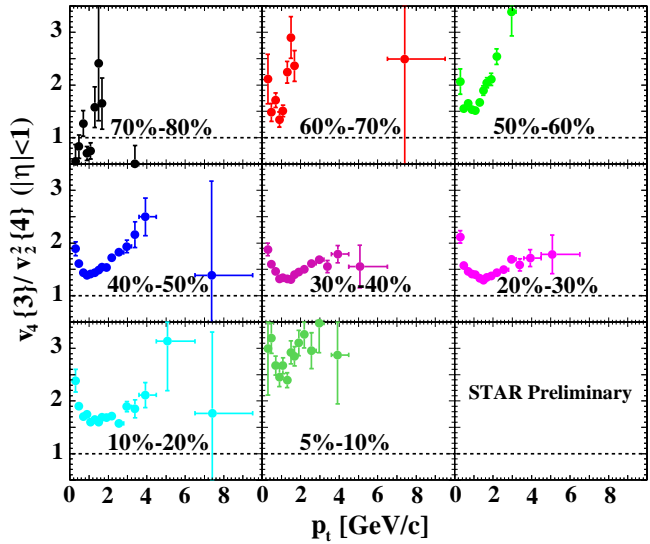}
}
\caption{Centrality dependence of $v_4\{3\}/v_2^2\{4\}$
in Au+Au collisions at $\sqrt{s_{NN}} =$ 200 GeV. 
From~\cite{TANG:2006xx}}
\label{fig:ratiov2v4ptcen} 
\end{figure}
A better comparison is the transverse momentum dependence of 
 $v_4/v_2^2$  shown in Fig.~\ref{fig:ratiov2v4ptcen} 
for eight centralities. 
Below \pt = 1.5 GeV/c the transverse momentum dependence is in agreement 
with hydro expectation. However for each of the centralities the minimum 
of $v_4/v_2^2$ is still more than a factor of two larger than the asymptotic 
ideal hydro value. 
Measurements of the energy dependence of this ratio, particularly at 
an order of magnitude higher beam energy at the LHC, should provide 
insight into the dynamics driving this ratio~\cite{Bhalerao:2005mm}.

\section{Conclusions}
\label{sec:conclusion}
At RHIC the observed large 
elliptic flow provides compelling evidence 
for strongly interacting matter which, in addition, appears to behave 
like an almost ideal fluid~\cite{RHIC_Discoveries}. 
At low \pt\ the ratio $v_4/v_2^2$, exhibits the transverse momentum 
dependence expected for an ideal fluid while, as expected, 
deviating at higher \pt. 
However, the magnitude of $v_4/v_2^2$ is still more than a factor of 
two larger than the asymptotic ideal hydro value.
At the LHC the expected increase in multiplicity together with 
the expected increase in anisotropic flow will allow for a detailed 
measurement of the $v_2$ and higher harmonics~\cite{ALICEppr}. 
These measurements are expected to quickly provide important 
confirmations, or perhaps new insights to our current 
understanding of the EoS of QCD matter.


\end{document}